
%
%
%
%

%
%
\catcode`@=11
%
%

\def\b@lank{ }

\newif\if@simboli
\newif\if@riferimenti
\newif\if@bozze

\def\bozze{\@bozzetrue\font\tt@bozze=cmtt8}
\def\og@gi{\number\day\space\ifcase\month\or
   gennaio\or febbraio\or marzo\or aprile\or maggio\or giugno\or
   luglio\or agosto\or settembre\or ottobre\or novembre\or dicembre\fi
   \space\number\year}
\newcount\min@uti
\newcount\or@a
\newcount\ausil@iario
\min@uti=\number\time
\or@a=\number\time
\divide\or@a by 60
\ausil@iario=-\number\or@a
\multiply\ausil@iario by 60
\advance\min@uti by \number\ausil@iario
\def\ora@esecuzione{\the\or@a:\the\min@uti}  
\def\makefootline{\baselineskip=24pt\line{\the\footline}
    \if@bozze\vskip-10pt\tt@bozze\noindent\og@gi, ore \ora@esecuzione.\fi}

\newwrite\file@simboli
\def\simboli{
    \immediate\write16{ !!! Genera il file \jobname.SMB }
    \@simbolitrue\immediate\openout\file@simboli=\jobname.smb}

\newwrite\file@ausiliario
\def\riferimentifuturi{
    \immediate\write16{ !!! Genera il file \jobname.AUX }
    \@riferimentitrue\openin1 \jobname.aux
    \ifeof1\relax\else\closein1\relax\input\jobname.aux\fi
    \immediate\openout\file@ausiliario=\jobname.aux}

\newcount\eq@num\global\eq@num=0
\newcount\sect@num\global\sect@num=0

\newif\if@ndoppia
\def\numerazionedoppia{\@ndoppiatrue\gdef\la@sezionecorrente{\the\sect@num}}

\def\se@indefinito#1{\expandafter\ifx\csname#1\endcsname\relax}
\def\spo@glia#1>{} 

\newif\if@primasezione
\@primasezionetrue

\def\s@ection#1\par{\immediate
    \write16{#1}\if@primasezione\global\@primasezionefalse\else\goodbreak
    \vskip\spaziosoprasez\fi\noindent
    {\bf#1}\nobreak\vskip\spaziosottosez\nobreak\noindent}
%

\def\sezpreset#1{\global\sect@num=#1
    \immediate\write16{ !!! sez-preset = #1 }   }

\def\spaziosoprasez{50pt plus 60pt}
\def\spaziosottosez{15pt}

\def\sref#1{\se@indefinito{@s@#1}\immediate\write16{ ??? \string\sref{#1}
    non definita !!!}
    \expandafter\xdef\csname @s@#1\endcsname{??}\fi\csname @s@#1\endcsname}

\def\autosez#1#2\par{
    \global\advance\sect@num by 1\if@ndoppia\global\eq@num=0\fi
    \xdef\la@sezionecorrente{\the\sect@num}
    \def\usa@getta{1}\se@indefinito{@s@#1}\def\usa@getta{2}\fi
    \expandafter\ifx\csname @s@#1\endcsname\la@sezionecorrente\def
    \usa@getta{2}\fi
    \ifodd\usa@getta\immediate\write16
      { ??? possibili riferimenti errati a \string\sref{#1} !!!}\fi
    \expandafter\xdef\csname @s@#1\endcsname{\la@sezionecorrente}
    \immediate\write16{\la@sezionecorrente. #2}
    \if@simboli
      \immediate\write\file@simboli{ }\immediate\write\file@simboli{ }
      \immediate\write\file@simboli{  Sezione
                                  \la@sezionecorrente :   sref.   #1}
      \immediate\write\file@simboli{ } \fi
    \if@riferimenti
      \immediate\write\file@ausiliario{\string\expandafter\string\edef
      \string\csname\b@lank @s@#1\string\endcsname{\la@sezionecorrente}}\fi
    \goodbreak\vskip\spaziosoprasez
    \noindent\if@bozze\llap{\tt@bozze#1\ }\fi
      {\bf\the\sect@num.\quad #2}\par\nobreak\vskip\spaziosottosez
    \nobreak\noindent}

\def\semiautosez#1#2\par{
    \gdef\la@sezionecorrente{#1}\if@ndoppia\global\eq@num=0\fi
    \if@simboli
      \immediate\write\file@simboli{ }\immediate\write\file@simboli{ }
      \immediate\write\file@simboli{  Sezione ** : sref.
          \expandafter\spo@glia\meaning\la@sezionecorrente}
      \immediate\write\file@simboli{ }\fi
    \s@ection#2\par}


\def\eqpreset#1{\global\eq@num=#1
     \immediate\write16{ !!! eq-preset = #1 }     }

\def\eqref#1{\se@indefinito{@eq@#1}
    \immediate\write16{ ??? \string\eqref{#1} non definita !!!}
    \expandafter\xdef\csname @eq@#1\endcsname{??}
    \fi\csname @eq@#1\endcsname}

\def\eqlabel#1{\global\advance\eq@num by 1
    \if@ndoppia\xdef\il@numero{\la@sezionecorrente.\the\eq@num}
       \else\xdef\il@numero{\the\eq@num}\fi
    \def\usa@getta{1}\se@indefinito{@eq@#1}\def\usa@getta{2}\fi
    \expandafter\ifx\csname @eq@#1\endcsname\il@numero\def\usa@getta{2}\fi
    \ifodd\usa@getta\immediate\write16
       { ??? possibili riferimenti errati a \string\eqref{#1} !!!}\fi
    \expandafter\xdef\csname @eq@#1\endcsname{\il@numero}
    \if@ndoppia
       \def\usa@getta{\expandafter\spo@glia\meaning
       \la@sezionecorrente.\the\eq@num}
       \else\def\usa@getta{\the\eq@num}\fi
    \if@simboli
       \immediate\write\file@simboli{  Equazione
            \usa@getta :  eqref.   #1}\fi
    \if@riferimenti
       \immediate\write\file@ausiliario{\string\expandafter\string\edef
       \string\csname\b@lank @eq@#1\string\endcsname{\usa@getta}}\fi}

\def\autoreqno#1{\eqlabel{#1}\eqno(\csname @eq@#1\endcsname)
       \if@bozze\rlap{\tt@bozze\ #1}\fi}
\def\autoleqno#1{\eqlabel{#1}\leqno\if@bozze\llap{\tt@bozze#1\ }
       \fi(\csname @eq@#1\endcsname)}

\def\numeriadestra{\let\autoeqno=\autoreqno}
\def\numaeriasinistra{\let\autoeqno=\autoleqno}
\numeriadestra

\newcount\cit@num\global\cit@num=0

\newwrite\file@bibliografia
\newif\if@bibliografia
\@bibliografiafalse

\def\lp@cite{[}
\def\rp@cite{]}
\def\trap@cite#1{\lp@cite #1\rp@cite}
\def\lp@bibl{[}
\def\rp@bibl{]}
\def\trap@bibl#1{\lp@bibl #1\rp@bibl}

\def\refe@renza#1{\if@bibliografia\immediate        
    \write\file@bibliografia{
    \string\item{\trap@bibl{\cref{#1}}}\string
    \bibl@ref{#1}\string\bibl@skip}\fi}

\def\ref@ridefinita#1{\if@bibliografia\immediate\write\file@bibliografia{
    \string\item{?? \trap@bibl{\cref{#1}}} ??? tentativo di ridefinire la
      citazione #1 !!! \string\bibl@skip}\fi}

\def\bibl@ref#1{\se@indefinito{@ref@#1}\immediate
    \write16{ ??? biblitem #1 indefinito !!!}\expandafter\xdef
    \csname @ref@#1\endcsname{ ??}\fi\csname @ref@#1\endcsname}

\def\c@label#1{\global\advance\cit@num by 1\xdef            
   \la@citazione{\the\cit@num}\expandafter
   \xdef\csname @c@#1\endcsname{\la@citazione}}

\def\bibl@skip{\vskip 0truept}


\def\stileincite#1#2{\global\def\lp@cite{#1}\global
    \def\rp@cite{#2}}
\def\stileinbibl#1#2{\global\def\lp@bibl{#1}\global
    \def\rp@bibl{#2}}

\def\citpreset#1{\global\cit@num=#1
    \immediate\write16{ !!! cit-preset = #1 }    }

\def\autobibliografia{\global\@bibliografiatrue\immediate
    \write16{ !!! Genera il file \jobname.BIB}\immediate
    \openout\file@bibliografia=\jobname.bib}

\def\cref#1{\se@indefinito                  
   {@c@#1}\c@label{#1}\refe@renza{#1}\fi\csname @c@#1\endcsname}

\def\cite#1{\trap@cite{\cref{#1}}}                  
\def\ccite#1#2{\trap@cite{\cref{#1},\cref{#2}}}     
\def\ncite#1#2{\trap@cite{\cref{#1}--\cref{#2}}}    
\def\upcite#1{$^{\,\trap@cite{\cref{#1}}}$}               
\def\upccite#1#2{$^{\,\trap@cite{\cref{#1},\cref{#2}}}$}  
\def\upncite#1#2{$^{\,\trap@cite{\cref{#1}-\cref{#2}}}$}  

\def\clabel#1{\se@indefinito{@c@#1}\c@label           
    {#1}\refe@renza{#1}\else\c@label{#1}\ref@ridefinita{#1}\fi}

\def\biblskip#1{\def\bibl@skip{\vskip #1}}           

\def\insertbibliografia{\if@bibliografia             
    \immediate\write\file@bibliografia{ }
    \immediate\closeout\file@bibliografia
    \catcode`@=11\input\jobname.bib\catcode`@=12\fi}


\def\commento#1{\relax}
\def\biblitem#1#2\par{\expandafter\xdef\csname @ref@#1\endcsname{#2}}


\catcode`@=12

\magnification=1200
\hsize=15truecm
\vsize=23truecm
\baselineskip 20 truept
\voffset=-0.5 truecm
\parindent=1cm
\overfullrule=0pt

\autobibliografia
\def\au{\autoeqno}
\def\e#1{(\eqref{#1})}

\noindent The $N=1$, $D=10$ SUGRA--SYM theory
is known to be plagued by an ABBJ gauge and Lorentz anomaly
which, for the gauge groups $SO(32)$ or $E_8\times E_8$, cancels
via the Green--Schwarz anomaly cancellation mechanism;
the harmlessness of this anomaly is reflected also in the
corresponding superstring theories \cite{GS}. It is, however,
immediately seen that the appearance of this gauge--Lorentz
anomaly ${\cal A}_G$ implies also the appearance of a (local)
supersymmetry anomaly ${\cal A}_S$. If we call $\Gamma$ the
effective action of the SUGRA--SYM theory and $\Omega_S(\Omega_G)$
the nihilpotent BRST operator associated to supersymmetry
transformations (gauge and Lorentz transformations) we have,
apart from $o(\hbar^2)$ terms,
$$
\Omega_G\Gamma = {\cal A}_G\au{1}
$$
where ${\cal A}_G$ is the usual ABBJ gauge--Lorentz anomaly
\ccite{ABBJ}{GSBOOK}. This anomaly is clearly not invariant under
supersymmetry transformations and so
$$
0\not = \Omega_S {\cal A}_G = \Omega_S\Omega_G\Gamma=-
\Omega_G(\Omega_S\Gamma)\au{2}
$$
meaning that
$$
\Omega_S\Gamma = {\cal A}_S\not =0,\au{3}
$$
i.e. $\Gamma$ suffers also a supersymmetry anomaly.
Eqs.\e{1}-\e{3} define the following coupled cohomology problem
$$\eqalign{
\Omega_G  {\cal A}_G &= 0\cr
\Omega_G {\cal A}_S + \Omega_S {\cal A}_G&=0\cr
\Omega_S {\cal A}_S &= 0\cr}\au{4}
$$
which, for the known ABBJ anomaly ${\cal A}_G$, can be considered
as a system of equations for the unknown supersymmetry anomaly
${\cal A}_S$, the ``supersymmetric partner'' of ${\cal A}_G$.
An analogous cohomology problem can be defined also for supersymmetric
chiral Yang--Mills theories in $D=4$ and 6 dimensions. Explicit solutions
for ${\cal A}_S$ satisfying \e{4} have been found for gauge anomalies
in $D=4,6$ \cite{WT} in the case of global and local supersymmetry and
for the Lorentz anomaly in pure supergravity theories in $D=6,10$
\cite{WT2} due to a remarkable property called ``Weil triviality''
(see below) in those references. However, in the case of the coupled
SUGRA--SYM theory in $D=10$, i.e. the theory under investigation in this
letter, it is difficult to prove Weil triviality and a general solution
of \e{4} has not yet been found.

As is well known, for the gauge groups $SO(32)$ and $E_8\times E_8$,
there exists a local counterterm $\Delta$ (see below) which
trivializes the cocyle ${\cal A}_G$
$$
{\cal A}_G=\Omega_G\Delta.\au{5}
$$
The information contained in \e{5}, together with the cohomology eqs. \e{4},
generally speaking, do not allow one to conclude that also the
supersymmetry anomaly trivializes.
In this letter we want a) solve \e{4} for these particular gauge groups
giving an explicit expression for ${\cal A}_S$, b) show that ${\cal A}_S$
is a trivial cocycle of $\Omega_S$ and c) show that ${\cal A}_S$ gets
trivialized by the same $\Delta$ appearing in \e{5}.
$$
{\cal A}_S=\Omega_S\Delta.\au{6}
$$
We would like to notice that, among all cases solved in the literature until
now, only
in the case of the SYM theory with global SUSY in $D=4$ dimensions
${\cal A}_S$ is known to be a trivial cocyle of $\Omega_S$
\cite{WT} (in that case,
this holds true even if ${\cal A}_G$ is non trivial).
To summarize our result: the Green--Schwarz counterterm cancels also
the supersymmetry anomaly.

Our notation is as follows. We work in a ten-dimensional superspace
span\-ned by the coordinates $z^M=(x^m,\vartheta^\mu)$ where
$x^m$ $(m=0,1,...,9)$ are the ordinary space--time coordinates and
$\vartheta^\mu$ $(\mu=1,...,16)$ are Grassmann variables. We introduce the
supervielbein one-forms $E^A=dz^M E_M{}^A(z)$ where $A=\{a,\alpha\}$
$(a=0,1,..,9;\  \alpha=1,...,16)$ is a flat index (letters from the
beginning of the alphabet represent flat indices, letters from the
middle of the alphabet onwards represent curved indices: small latin
letters indicate vectorial indices, small greek letters indicate spinorial
indices and capital letters denote both of them). $p$-superforms can be
decomposed along the vielbein basis or along the coordinate basis,
$$
\phi = {1\over p!} E^{A_1}\cdots E^{A_p} \phi_{A_p\cdots A_1}={1\over p!}
dz^{M_1}\cdots dz^{M_p}\phi_{M_p\cdots M_1}.\au{6a}
$$
We denote the Lorentz-valued super spin connection one-form by
$\omega_a{}^b=dz^M \omega_{Ma}{}^b=E^C\omega_{Ca}{}^b$
($\omega_{ab}=-\omega_{ba}$)
and the corresponding Lorentz--curvature two-superform by
$R_a{}^b=d\omega_a{}^b + \omega_a{}^c \omega_c{}^b={1\over 2}
E^DE^CR_{CDa}{}^b$. $d$ indicates the differential in superspace.
We introduce a Yang--Mills connection one-superform $A=E^B A_B$ with
values in the Lie algebra of a group $H$ ($SO(32)$ or $E_8\times E_8$
in the following) and the Lie--algebra valued curvature two-superform
$F=dA+AA={1\over 2} E^B E^C F_{CB}$. To construct the relevant nihilpotent
BRST operators we introduce the ghost fields $\xi^M(z)$ for
superdiffeomorphisms, a Lie algebra valued ghost field $C(z)$ for
gauge transformations and an $SO(10)$ Lorentz-valued ghost field
$u_a{}^b(z)$ for local Lorentz transformations. Local supersymmetry
transformations are represented by the superdiffeomorphisms. The BRST
transformations are the following:

\noindent
Superdiffeomorphisms:
$$\eqalign{
\delta_S\xi^M &= \xi^L\partial_L\xi^M\cr
\delta_S C &= \xi^L \partial_L C\cr
\delta_S u_a{}^b &= \xi^L \partial_L u_a{}^b\cr
\delta_S\phi &= L_\xi \phi \equiv (d\,i_\xi-i_\xi d)\phi\cr}
\au{7}
$$
\smallskip
\noindent
Gauge transformations:
$$\eqalign{
\delta_H C &=-CC\cr
\delta_H A &= - dC-AC-CA\cr
\delta_H u_a{}^b &= \delta_H \xi^M =
\delta_H \omega_a{}^b = 0\cr}\au{8}
$$
\smallskip
\noindent
Local Lorentz transformations:
$$\eqalign{
\delta_L u_a{}^b &=- u_a{}^c u_c{}^b\cr
\delta_L \omega_a{}^b &= - du_a{}^b - \omega_a{}^c
u_c{}^b-u_a{}^c\omega_c{}^b\cr
\delta_L C &= \delta_L \xi^M = \delta_L A=0\cr
\delta_L V_{A_1\cdots A_n} &= - u_{A_1}{}^B
V_{BA_2\cdots A_n}-\ ({\rm permutations})\cr}\au{9}
$$
$L_\xi$ indicates the Lie derivative along the vector field
$\xi=\xi^M\partial_M$, the interior product with a superform $\phi$
is defined as $i_\xi\phi={1\over (p-1)!} \xi^{M_1}
dz^{M_2}\cdots dz^{M_p} \phi_{M_p\cdots M_1}$,
$u_\alpha{}^\beta = {1\over 4} (\Gamma_{ab})_\alpha{}^\beta
u^{ab}$, $u_a{}^\alpha=u_\alpha{}^a =0$, and $V_{A_1\cdots A_n}$
in \e{9} indicates a Lorentz--tensor. We note the useful relation
$$
\delta_S(i_\xi \phi) = {1\over 2}
(d\,i_\xi i_\xi - i_\xi i_\xi d)\phi \au{9a}
$$
where $\phi$ is a superform with zero ghost number.

The superfields (or forms) of the theory constitute a graded algebra
if we define the grading $n_\psi$ of a $p$-form $\psi$ as $n_\psi=p+g_\psi$,
where $g_\psi$ is the ghost number of $\psi$, and associate a grading 1 to the
operators $d$ and $\delta$. $d$ and $\delta$ begin to operate on a
composite term from the right. There are the usual additional signs if
the fields carry spinorial indices \cite{WB}.

It is convenient to combine the transformations in \e{8} and \e{9} to
define BRST transformations associated to the ``total" gauge group
$G=H\otimes SO(10)$ through
$$
\delta_G \equiv \delta_H + \delta_L.\au{10}
$$
Then we can associate to $\delta_S$ and $\delta_G$ the corresponding
BRST operators $\Omega_S$  and $\Omega_G$, satisfying:
$\Omega^2_G = \Omega_S\Omega_G+\Omega_G\Omega_S=\Omega^2_S=0$. These
operators define then the coupled cohomology problem \e{4}. We try to
solve it, for $H=SO(32)$ or $E_8\times E_8$, according to a method
developed in \ccite{WT}{WT2} which represents a generalization of the
method for the determination of ABBJ anomalies, based on the extended
transgression formula (ETF) \cite{ETF}, to the supersymmetric case.

For the above gauge groups the relevant sixth-order polynomial in the
curvatures factorizes, as is well known \cite{GSBOOK}, into
$$
I_{12} = X_4 X_8 \au{11}
$$
with
$$\eqalign{
X_4 &= {1\over 30} Tr F^2 - tr R^2\cr
X_8 &= {1\over 24} Tr F^4 - {1\over 7200} (Tr F^2)^2 - {1\over 240} Tr
F^2 tr R^2 + {1\over 8} tr R^4 + {1\over 32} (tr R^2)\cr}\au{12}
$$
($Tr$ denotes the trace in the adjoint representation, $tr$ denotes the
trace in the fundamental representation).
We denote collectively ${\cal F}=(F,R)$, ${\cal A}=(A,\omega)$,
${\cal C}=(C, u_a{}^b)$. $X_4$ and $X_8$ are closed $G$-invariant
superforms. For the moment we pursue formally the ETF method. Define:
$$
\hat d = d+\delta_G\au{13}
$$
$$
\hat{\cal A} = {\cal A} + {\cal C}\au{14}
$$
$$
\hat{\cal F}=\hat d\hat{\cal A}+\hat{\cal A}\hat{\cal A}=
{\cal F}\au{15}
$$
$$
\hat{\cal F}_t = t(\hat d\hat{\cal A}+t\hat{\cal A}\hat{\cal A})
\au{16}
$$
where the real parameter $t$ runs from 0 to 1. Then one has the identities
$$\eqalign{
X_4 &= \hat d\, \left( 2\int^1_0 dt \ X_4 (\hat{\cal A},\hat{\cal F}_t)
\right)\equiv\hat d\, Q_3\cr
X_8 & = \hat d\, \left(4 \int^1_0 dt \ X_8 (\hat{\cal A},\hat{\cal F}_t)
\right) \equiv \hat d\, Q_7.\cr}\au{17}
$$
With the integrands in \e{17} we mean the expressions which are
obtained from $X_4$ and $X_8$ by substituting {\it one} of the curvatures
with $\hat{\cal A}$ and the remaining ones with $\hat{\cal F}_t$ and
symmetrizing then with unit weight (the explicit expressions of $Q_3$ and
$Q_7$ are, however, not relevant for our purposes). $Q_3$ and $Q_7$
can be decomposed in sectors with different ghost number $p$
$$\eqalign{
Q_3 &= \sum^3_{p=0} X_{p,3-p}\cr
Q_7 &= \sum^7_{p=0} X_{p,7-p}\cr}\au{18}
$$
\noindent
and \e{17} implies then the following identities (descent equations)
$$\eqalign{
d X_{0,3} &= X_4\cr
d X_{1,2} + \delta_G X_{0,3} &=0\cr
d X_{2,1} + \delta_G X_{1,2} &=0\cr}\au{19}
$$
and
$$\eqalign{
d X_{0,7} & = X_8\cr
d X_{1,6} + \delta_G X_{0,7} &=0\cr
d X_{2,5} + \delta_G X_{1,6} &=0.\cr}\au{20}
$$
Moreover, due to the fact that $X_4$ and $X_8$ are G-invariant and
closed we have
$$\eqalign{I_{12} &= X_4 X_8 =
{2\over3}\,X_4\,\hat d\,Q_7 + {1\over 3}\,\hat d\, Q_3X_8\cr
& = \hat d\, \left( {2\over 3}\,X_4 Q_7 + {1\over 3}\,Q_3 X_8 \right).
\cr}\au{21}
$$
If we were now in ordinary space the left hand side in \e{21}
would vanish, being a twelve form in ten dimensions, and the first non
trivial descent equation associated to \e{21} would identify
the ABBJ G-anomaly. In superspace, however, $I_{12}$ is non vanishing.
This problem can be bypassed if, in superspace, $I_{12}$ can be
written as the differential of a {\it $G$-invariant\/}
eleven-superform $Y_{11}$ (``Weil triviality" \cite{WT2})
$I_{12}=d Y_{11}$. As has been shown in \cite{WT2} such  a
$G$-invariant $Y_{11}$ would exist if one could impose on the Yang-Mills and
gravitational supercurvatures the constraints
$F_{\alpha\beta}=0$, $R_{\alpha\beta a}{}^b=0$. While the first constraint
is always available it is known that for the coupled SUGRA--SYM
theory $R_{\alpha\beta a}{}^b$ is intrinsically non vanishing, in the
sense that it can not be set to zero by any field redefinition
\ccite{HULL}{CL}. In the absence of such a constraint until now it was
not possible to prove Weil triviality  of $I_{12}$ for an arbitrary gauge
group $H$.

However, and this is the key observation of this letter, the theory at
hand contains a two-form potential $B$ among its physical fields with
associated $G$-invariant curvature $H$ defined as
$$\eqalign{
H &= dB + X_{0,3}\cr
dH &= X_4 = {1\over 30} Tr F^2-tr R^2\cr
\delta_G B &= - X_{1,2}\cr
\delta_G H &=0.
\cr}\au{22}
$$
The second crucial point is that a consistent exact solution of \e{22} in
{\it superspace\/} has been given in \cite{BPT}.
This means that all unphysical fields have been eliminated and that the
remaining physical fields are subjected to SUSY transformation rules which
represent correctly the algebra of local supersymmetry transformations, and
assure in particular the nihilpotency of the corresponding BRST operator.
Using \e{22} it is now easy to prove Weil triviality for $I_{12}$, in fact:
$$
I_{12} = X_4 X_8 = dH X_8 = d\,(H X_8) = \hat d\,(H X_8),\au{23}
$$
where the last step stems from the fact that $\delta_G(HX_8)=0$.
Now we can proceed along the lines of \cite{WT2}. \e{21} and \e{23} imply
$$
0=\hat d\, \left( {2\over 3}\,X_4Q_7 +
{1\over 3}\,Q_3X_8- H X_8\right) \equiv\hat d\, Q.\au{24}
$$
As above we decompose $Q$ in sectors with definite ghost number
$p$, $Q=\sum^{11}_{p=0}Q_{p,11-p}$, and \e{24}
implies then the descent equations:
$$
\eqalign{
dQ_{0,11} &= 0\cr
dQ_{1,10} + \delta_G Q_{0,11} &=0\cr
dQ_{2,9} + \delta_G Q_{1,10} &=0.\cr}\au{25}
$$
Applying $i_\xi$ twice to the first equation in \e{25} and once to
the second and recalling the definition of the Lie derivative
${\cal L}_\xi$ and \e{9a} we get:
$$\eqalign{
\delta_S(i_\xi Q_{0,11}) - {1\over 2} d\, i_\xi i_\xi Q_{0,11} &= 0\cr
d(i_\xi Q_{1,10}) - \delta_S Q_{1,10} + \delta_G (i_\xi Q_{0,11}) &=0\cr
d Q_{2,9} + \delta_G Q_{1,10} &=0.\cr}\au{26}
$$
Defining now the integral of a ten-superform $\psi_{10}$ over ordinary
space--time as
$$
\int\psi_{10} \equiv {1\over 10!}\,\int d^{10}x\,
\varepsilon^{m_1-m_{10}} \psi_{m_{10}\ldots m_1} (z)\Bigl\vert_{\vartheta=0}
\au{27}
$$
and noting that the integral of an exact ten-superform is zero,
$\int d \psi_9=0$, we can integrate eqs. \e{26} and obtain
a solution  of the coupled cohomology problem \e{4} with:
$$\eqalign{
{\cal A}_G &= \int Q_{1,10}\cr
{\cal A}_S &= - \int i_\xi Q_{0,11}.\cr}\au{28}
$$
It is easy to extract from \e{24}, using \e{18}, the components of $Q$
with ghost number 1 and 0:
$$
{\cal A}_G =\int \left( {2\over 3}\, X_4X_{1,6} + {1\over 3}\, X_{1,2}X_8
\right)\au{29}
$$
$$
{\cal A}_S = \int i_\xi \left( H X_8 - {1\over 3}\, X_{0,3} X_8 -
{2\over 3}\, X_4 X_{0,7}\right).\au{30}
$$
Eq. \e{29} is clearly the usual expression of the ABBJ $G$-anomaly
\cite{GSBOOK},
while eq. \e{30} represents the supersymmetry anomaly we searched for.
Using eqs. \e{19}, \e{20} and \e{22}, we can rewrite it as follows:
$$\eqalign{
{\cal A}_S & = \int i_\xi \left( d(BX_8)+{2\over 3}\, (X_{0,3} X_8-
X_4 X_{0,7})\right)\cr
& = \Omega_S \int \left(-B X_8 -{2\over 3}\,
X_{0,3} X_{0,7}\right)\cr
& \equiv \Omega_S \Delta\cr}\au{31}
$$
meaning that ${\cal A}_S$ is a trivial cocycle of $\Omega_S$. But it is
easy to recognize $\Delta$ as the famous Green--Schwarz counterterm which
cancels the $G$-anomaly ${\cal A}_G$ (use again eqs. \e{19},\e{20})
$$
{\cal A}_G = \Omega_G \Delta.\au{32}
$$
This means that if we redefine the effective action $\Gamma$ according to
$\tilde\Gamma=\Gamma-\Delta$ we obtain a $G$-invariant and
supersymmetric theory. This is what one expects due to the fact that the
anomaly ${\cal A}_S$ gets induced by the gauge anomaly, and one can argue that
the cancellation of the latter implies also the vanishing of the former.
Clearly our argument does not exclude the presence of other (non trivial)
supersymmetry cocycles which are not related to the ABBJ anomaly. This would,
however, spoil the quantum consistency of the theory.

As a last remark we point out that the absence of SUSY anomalies in the
$N=1$, $D=10$ SUGRA--SYM theory reflects the fact that also the related
superstring theories are not plagued by SUSY anomalies.

Recently, in the literature a string-five-brane duality conjecture has been
made \ccite{DUFF}{STRO}; it states that, in their critical space-time
dimensions $D=10$,
superstrings (extended objects with one spatial dimension) are dual to
super five-branes (extended objects with five spatial dimensions). In
particular, in \cite{PLEFKA} the gauge and Lorentz anomalies of the heterotic
five-brane
sigma model in an $N=1$, $D=10$ SUGRA-SYM background have been determined by
counting the chiral fermions and relying then on the index theorem. The net
result is that the fourth-order polynomial responsible for the gauge and
Lorentz anomalies of the sigma model
is given precisely by $X_8$ (see \e{12}). To cancel these
sigma model anomalies, instead of introducing a three-form $H_3$ satisfying
$dH_3=X_4$,
one postulates the existence of a ``dual'' SUGRA-SYM theory based on a seven
form $H_7$ \cite{PLEFKA} satisfying
$$
dH_7 = X_8.\au{33}
$$
Clearly, as is known, eq. \e{33} assures the cancellation of the ABBJ
anomalies also in the
``dual'' supergravity theory \cite{GATES}. In this case again a supersymmetric
partner ${\tilde{\cal A}}_S$ of
these anomalies arises, and our procedure to compute it and to show its
harmlessness could be repeated in a straightforward
way simply by interchanging the roles of $X_4$ and $X_8$ once a
consistent solution of \e{33} in {\it superspace\/} is obtained. But such a
solution is still missing and seems difficult to exist. A small inspect
into the Bianchi identity \e{33} in superspace reveals that,
to obtain a consistent solution, one has at least to give
up the rigid supersymmetry preserving constraint
$$
T_{\alpha\beta}{}^a = 2\Gamma^a_{\alpha\beta}\au{34}
$$
by introducing a 1050 irreducible representation of $SO(10)$,
$W_{c_1\ldots c_5}{}^a$, according to
$$
T_{\alpha\beta}{}^a = 2\Gamma^a_{\alpha\beta}+
(\Gamma^{c_1\ldots c_5})_{\alpha\beta}W_{c_1\ldots c_5}{}^a.\au{34a}
$$
In fact, as has been shown in ref. \cite{CL}, eq. \e{34} leads necessarily
to a {\it Lorentz and gauge-invariant\/} $H_7$ satisfying $dH_7=0$.
Even with \e{34a}, it is by no means obvious that
one can solve the Bianchi identity  \e{33} consistently; on the other hand,
to our knowledge no supergravity
theory is known in which \e{34} is violated. This points in the direction of
a conflict between the heterotic five-brane and supersymmetry.
\bigskip
\noindent{\it Acknowledgements}\par
\vskip0.3truecm\noindent
The authors would like to thank Mario Tonin for useful discussions.
\vskip 1.4truecm
\noindent{\it References}\par
\vskip0.3truecm\noindent
%
\biblitem{GS} M.~B.~Green and J.~H.~Schwarz, {\it Phys. Lett.} {\bf 149B}
(1984) 117\par
\biblitem{GATES} A.~Salam and E.~Sezgin, {\it Phys. Scripta}
{\bf32} (1985) 283;
S.~J.~Gates and H.~Nishino, {\it Phys. Lett.} {\bf 157B}
(1985) 157\par
\biblitem{GS2} M.~B.~Green and J.~H.~Schwarz, {\it Phys. Lett.} {\bf 151B}
(1985) 21\par
%
\biblitem{GSBOOK} M.~B.~Green, J.~H.~Schwarz and E.~Witten, {\it Superstring
Theory}, Cambridge University Press, Cambridge (1987)\par
%
\biblitem{ABBJ} S.~J.~Adler, {\it Phys. Rev.} {\bf177} (1969) 2426;
J.~S.~Bell and R.~Jackiw, {\it Nucl. Phys.} {\bf A60} (1969) 47;
L.~Bonora and P.~Pasti, {\it Phys. Lett.} {\bf132B} (1983) 75\par
%
\biblitem{WT} L.~Bonora, P.~Pasti and M.~Tonin, {\it Phys. Lett.} {\bf 156B}
(1985) 341; {\it Nucl. Phys.} {\bf B261} (1985) 249; {\it Phys. Lett.} {\bf
167B} (1986) 191\par
%
\biblitem{WT2} L.~Bonora, P.~Pasti and M.~Tonin, {\it Nucl. Phys.} {\bf B286}
(1987) 150\par
%
\biblitem{WB} J.~Wess and J.~Bagger, {\it Supersymmetry and Supergravity},
Princeton University Press, Princeton (1983)\par
%
\biblitem{BPT} L.~Bonora, P.~Pasti and M.~Tonin, {\it Phys. Lett.} {\bf 188B}
(1987) 335;
L.~Bonora, M.~Bregola, K.~Lechner, P.~Pasti and M.~Tonin, {\it
Int. J. Mod. Phys.} {\bf A5} (1990) 461;
R.~D'Auria and P.~Fr\'e, {\it Phys. Lett.} {\bf 200B} (1988) 63;
R.~D'Auria, P.~Fr\'e, M.~Raciti and F.~Riva {\it Int. J. Mod. Phys.}
{\bf A3} (1988) 953; L.~Castellani, R.~D'Auria and P.~Fr\'e, {\it Phys. Lett.}
{\bf 196B} (1987) 349\par
%
\biblitem{ETF} L.~Bonora and P.~Cotta--Ramusino, {\it Phys. Lett.} {\bf 107B}
 (1981) 87;
 R.~Stora, {\it Carg\`ese Lectures} (1983);
 B.~Zumino, {\it Les Houches Lectures} (1983);
 B.~Zumino, Y.~S.~Wu and Z.~Zee, {\it Nucl. Phys.} {\bf B239} (1984) 477\par
%
\biblitem{STRO} A.~Strominger, {\it Nucl. Phys.} {\bf B343} (1990) 167\par
\biblitem{DUFF} M.~J.~Duff, {\it Class. Quantum Grav.} {\bf 5} (1988) 189\par
\biblitem{PLEFKA} J.~A.~Dixon, M.~J.~Duff and J.~C.~Plefka, {\it Phys. Rev.
Lett.} {\bf 69} (1992) 3009\par
%
\biblitem{CL} A.~Candiello and K.~Lechner, {\it Nucl. Phys.} {\bf B}412 (1994)
479\par
%
\biblitem{CLUN} A.~Candiello and K.~Lechner, {\it unpublished}\par
\biblitem{HULL} C.~M.~Hull, {\it Phys. Lett.} {\bf 167B} (1986) 51\par
\insertbibliografia
\vfill
\eject

\vsize=25truecm
\baselineskip 16truept
\nopagenumbers
\rightline{DFPD/94/TH/01}
\rightline{hep-th/9404095}
\vskip 1truecm
\centerline{{\bf THE SUPERSYMMETRIC VERSION OF THE}}
\smallskip
\centerline{{\bf GREEN--SCHWARZ ANOMALY CANCELLATION
MECHANISM}\footnote{*}{Supported
in part by M.P.I. This work is carried out in the framework of
the European Community Programme ``Gauge Theories, Applied Supersymmetry
and Quantum Gravity'' with a financial contribution under contract SC1-CT92
-D789.}
}
\vskip 1truecm
\centerline{\bf Antonio Candiello and Kurt Lechner}
\vskip 1truecm
\centerline{\sl Dipartimento di Fisica, Universit\`a di Padova}
\smallskip \centerline{\sl and}
\centerline{\sl Istituto Nazionale di Fisica Nucleare, Sezione di Padova}
\smallskip\centerline{Italy}
\vskip 2truecm

\centerline{\bf Abstract}
\vskip 1truecm
\noindent
The $N=1$, $D=10$ Supergravity--Super--Yang--Mills (SUGRA-SYM)
theory is plagued by
ABBJ gauge and Lorentz anomalies which are cancelled via the
Green-Schwarz anomaly cancellation mechanism. Due to the fact that the ABBJ
anomalies are not invariant under supersymmetry (SUSY) transformations
one concludes that the theory is plagued also by a SUSY anomaly. For
the gauge groups $SO(32)$ and $E_8\times E_8$ we compute this SUSY
anomaly, by solving a coupled cohomology problem, and we show that
it can be cancelled by subtracting from the action the known Green--Schwarz
counterterm, the same which cancels also the ABBJ anomaly, the
expected result. Finally, we argue that the corresponding mechanism does
not apply in the dual SUGRA-SYM, related to the heterotic five-brane.
\vfill\eject
\bye